\documentclass[10pt,conf]{IEEEtran}
\usepackage{geometry}
 \usepackage{bbold}
\usepackage{ifpdf}
\usepackage{graphicx}
\usepackage{subcaption}%
\usepackage[cmex10]{amsmath}
\usepackage{amssymb}
\usepackage{verbatim}
\usepackage{algorithm}
\usepackage{array}
\usepackage{latexsym}
\usepackage{color}
\usepackage{textgreek}
\usepackage{subscript}
\usepackage{rotating}
\usepackage{booktabs,multirow}
\usepackage{mdframed}	
\usepackage{tabularx}
\usepackage{amsmath}
\usepackage{makecell}
\usepackage{tabto}
\usepackage{algpseudocode}

\usepackage{mathrsfs}
\usepackage{url}
\usepackage{cite}
\usepackage{listings}
\usepackage{array}
\usepackage[]{xcolor}

\usepackage{microtype}

\newcommand{\nref}[1]{\textcolor{red}{[{\bf NEED CITATION}]}}
\algdef{SE}{Begin}{End}{\textbf{begin}}{\textbf{end}}
\algnewcommand\algorithmicforeach{\textbf{for each:}}
\algnewcommand\ForEach{\item[ \algorithmicforeach]}

\begin{document}
\setlength{\columnsep}{0.2 in}
\def\BibTeX{{\rm B\kern-.05em{\sc i\kern-.025em b}\kern-.08em T\kern-.1667em\lower.7ex\hbox{E}\kern-.125emX}}

\title{Generative Adversarial Network-Driven Detection of Adversarial Tasks in Mobile Crowdsensing }
\author{
	Zhiyan Chen~\IEEEmembership{Student Member,~IEEE} and Burak Kantarci~\IEEEmembership{Senior Member,~IEEE}
       
        \IEEEcompsocitemizethanks{
		   \IEEEcompsocthanksitem The authors are with the School of Electrical Engineering and Computer Science, University of Ottawa, Ottawa, ON, Canada.  emails: \{zchen241, burak.kantarci\}@uottawa.ca

          }
}

\maketitle
\thispagestyle{empty}
\pagestyle{empty}

\begin{abstract}
Mobile Crowdsensing systems are vulnerable to various attacks as they build on non-dedicated and ubiquitous properties. Machine learning (ML)-based approaches are widely investigated to build attack detection systems and ensure MCS systems security. However, adversaries that aim to clog the sensing front-end and MCS back-end leverage intelligent techniques, which are challenging for MCS platform and service providers to develop appropriate detection frameworks against these attacks. Generative Adversarial Networks (GANs) have been applied to generate synthetic samples, that are extremely similar to the real ones, deceiving classifiers such that the synthetic samples are indistinguishable from the originals. Previous works suggest that GAN-based attacks exhibit more crucial devastation than empirically designed attack samples, and result in low detection rate at the MCS platform. With this in mind, this paper aims to detect intelligently designed illegitimate sensing service requests by integrating a GAN-based model. To this end, we propose a two-level cascading classifier that combines the GAN discriminator with a binary classifier to prevent adversarial fake tasks. Through simulations, we compare our results to a single-level binary classifier, and the numeric results show that proposed approach raises Adversarial Attack Detection Rate (AADR), from $0\%$ to $97.5\%$ by KNN/NB, from $45.9\%$ to $100\%$ by Decision Tree. Meanwhile, with two-levels classifiers, Original Attack Detection Rate (OADR) improves for the three binary classifiers, with comparison, such as NB from $26.1\%$ to $61.5\%$.
\end{abstract}
\begin{IEEEkeywords}
\setlength{\columnsep}{0.2 in}
	Mobile Crowdsensing, Internet of Things, Machine Learning, GAN, Adversarial Detection.\end{IEEEkeywords}

\IEEEpeerreviewmaketitle

\section{Introduction}
Mobile Crowdsensing (MCS) enables individuals to provide sensing services via embedded sensors that are built in various smart equipment such as mobile phones, connected vehicles, dedicated sensors and wearables. As it is an effective business model for data acquisition, integration of MCS-based services with various applications have been explored such as smart transportation systems, smart grids, public safety, waste management, smart mobility and smart living \cite{Capponi.2019}. On the other hand, open, ubiquitous and non-dedicated deployment of sensing services in MCS systems remain vulnerabilities against data poisoning, fake sensing service requests, malware, and eavesdropping \cite{chen2021blockchain, Liu.2020, smahi2020blockchainized}.

Sensing service requests that aim at clogging the sensing and computing resources are known as fake task submission attacks \cite{zhang2019ai}. 
When adversaries do not possess sufficient resources to design advanced and intelligent models to generate adversarial tasks, the features that characterize the fake tasks are determined empirically and following certain distributions. 
We refer to these type of fake task attacks as original fake tasks. As they do not rely on intelligent design, the success of detecting these type of tasks is quite high as shown in the previous studies \cite{zhang2019ai}.

\begin{figure}[ht]
    \centering
    \includegraphics[width = 0.4\textwidth, trim=0.2cm 0.2cm 0.3cm 0.3cm,clip]{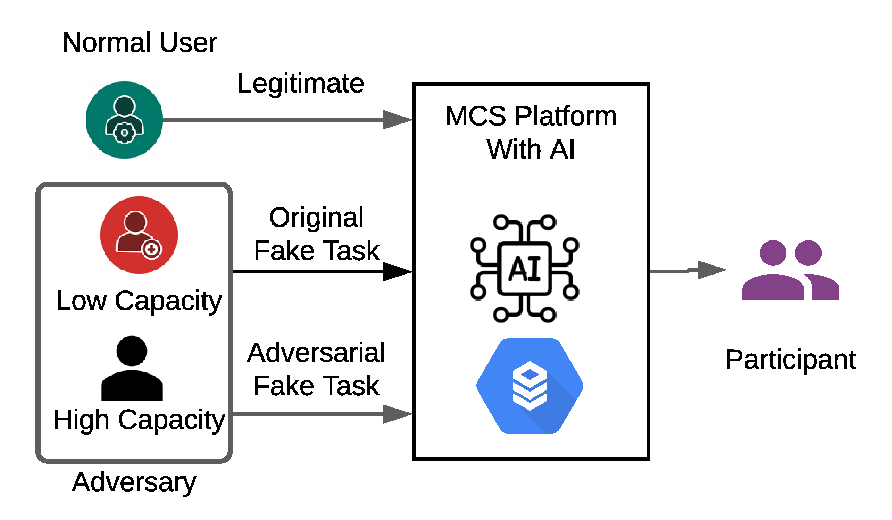}
    \caption{ Adversarial fake tasks, original fake tasks and benign tasks are submitted to MCS platform}
    \label{fig:MCS_Fake_Leg}
\end{figure}

Adversaries with high capacity generate fake tasks by leveraging adversarial machine learning. We refer to this type of fake tasks as adversarial fake tasks. Fig. \ref{fig:MCS_Fake_Leg} illustrates a snapshot of MCS platform that receives two types of tasks: legitimate tasks by normal users, and fake tasks (i.e., fake sensing service requests) by adversaries with either high or low capacity. Without training the fake task classifier (to pass legitimate tasks) in the MCS platform with adversarial samples, it is more challenging to detect the fake tasks that are generated by adversarial machine learning models. In our previous study \cite{chencisose2022}, we presented the potential of the adversaries with high capacity to generate fake tasks to clog the MCS system resources. Specifically, adversaries with high capacity deploy Generative Adversarial Network (GAN) \cite{creswell2018generative} to create fake tasks via training the generator model and discriminator model in GAN. In the previous study, the evaluation results show $0$ detection accuracy for generated fake tasks by GAN. Therefore, adversarial fake tasks (i.e., fake tasks generated through adversarial machine learning models) result in higher damages (i.e., low detection rate) to MCS systems when compared to fake tasks generated by adversaries with low capability.

\begin{figure}[ht]
\centering
    \includegraphics[width = 0.45\textwidth, trim=0.9cm 10.4cm 1.5cm 10.7cm,clip]{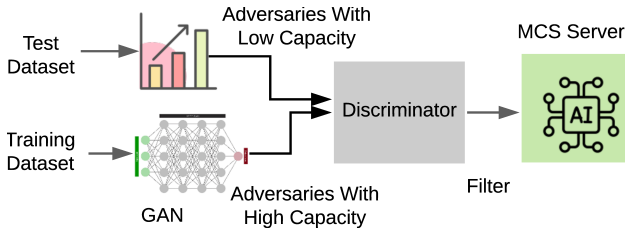}
    \caption{Discriminator and two types of adversary injections}
    \label{fig:Two_type_adver_new}
\end{figure}

In this paper, we extend our previous work in \cite{chencisose2022} and adopt the same approach to generate adversarial fake tasks. However, we aim to boost the detection performance for the adversarial fake tasks to ensure security of MCS systems. We propose a cascaded framework that consists of two-level classification: (1) level 1 includes a trained discriminator in GAN, (2) level 2 consists of a legacy binary classifier (e.g., KNN, Decision Tree (DT), Naive Bayes (NB)). The discriminator is responsible for recognizing adversarial attacks before feeding to level 2 classifiers. Fig.\ \ref{fig:Two_type_adver_new} demonstrates empirically designed fake task attacks in the test dataset (by adversaries with low capacity) and adversarial fake task attacks (by adversaries with high capacity) are fed to the discriminator before sending to the MCS server. 

Fig. \ref{fig:MCS_system} demonstrates the proposed system architecture. A GAN model is trained to generate adversarial fake tasks. The adversarial fake tasks are combined with the original tasks in test dataset and form the mixed dataset. Then, the trained discriminator model decides whether a task is adversarial or original since the discriminator has been trained to distinguish synthetic samples from original samples during the GAN training procedure. Tasks predicted as real by the discriminator are fed into level 2 classifiers to estimate whether they are fake or legitimate. Numeric results show that Adversarial Attack Detection Rate (AADR) under the cascaded model is improved up to $97.5\%$ when KNN/NB is employed in the second step, and up to $100\%$ when a Decision Tree is employed in the second step. Moreover, Original Attack Detection Rate (OADR) increases dramatically, from $26.1\%$ to $61.5\%$ under NB.

The rest of the paper is organized as follows. Section \ref{sec:background} presents the state of the art in machine learning-backed security for MCS systems. In Section \ref{sec:AML}, we introduce GAN-based adversarial task detection methodology in detail. Section \ref{sec:results} presents the numerical results and evaluations regarding the presented adversarial model. Finally, the paper is concluded in Section \ref{sec:conclusion}. 

\section{Related Work and Motivation}
\label{sec:background}

As tackled by the previous studies and reviewed in \cite{chen2021blockchain}, MCS systems confront several security vulnerabilities, such as data poisoning, fake sensing data and user privacy information leaking. Various methods have been investigated to secure MCS systems and prevent from the threats against MCS platforms and participants. 
The study in \cite{zhu2020deep} uses Rome as the setting/terrain to acquire the data, and leverages a neural network-based solution to improve the quantity of vehicular crowd-sensed data.
Moreover, the work in \cite{prud2021poisoning} tackles data poisoning attacks on MCS platforms and the authors present the impact of data poisoning attacks on several machine learning classifiers when poisoning data are injected by the adversaries that are equipped with capabilities to build and run AI models. The study in \cite{li2020quick} proposes a Lightweight Low Rank and False Matrix Separation (LightLRFMS) scheme to accelerate the computation time to detect false sensing data at significantly higher accuracy levels in comparison to the other separation / decomposition techniques.

Generative Adversarial Networks (GANs) are used to generate synthetic samples by training two neural networks that compete with each other, namely a discriminator and a generator. Typically, a GAN is capable of generating samples to pursue data augmentation the training dataset, which is known as GAN-based oversampling. 
For instance, the study in \cite{bouzeraib2020multi} demonstrates a GAN-based augmentation approach to tackle the challenge of limited attack samples in intelligent transportation systems. 
On the basis of the research in \cite{barni2020cnn} and other similar studies in the area, it is known that GANs can generate synthetic pictures which are not visually distinguishable from their natural samples. GANs can be utilized in various other use cases. 
For instance, the authors in \cite{elazab2020gp} present a case study in predicting growth of glioma via a innovative approach integrating Gp-GAN.
However, adversaries could also use GANs to deceive the machine learning classifiers such as those deployed in MCS platforms. Without proper training with synthetic samples, traditional machine learning models are not capable to distinguish the generated (i.e, also known as adversarial, synthetic) samples from real ones. Therefore, an effective method is required to identify the adversarial patterns alongside the attacks that are not the output of generative models but still aim at attacking the system.

As studied in \cite{chencisose2022}, when adversaries deploy GANs to generate fake tasks injected into an MCS platform that already runs ML-based fake task detection. In the absence of adversarial training, the detection rate is $0$ for these adversarial attacks. The synthetic fake tasks pose more crucial challenges to the MCS platform than original fake tasks. With this in mind, this paper builds on the work in \cite{chencisose2022}, with the motivation of boosting the detection rate of the adversarial fake task attacks to secure the MCS systems. %

\begin{figure}[ht]
    \centering
    \includegraphics[width = 0.9\linewidth, trim=0.6cm 1cm 0.6cm 0.6cm,clip]{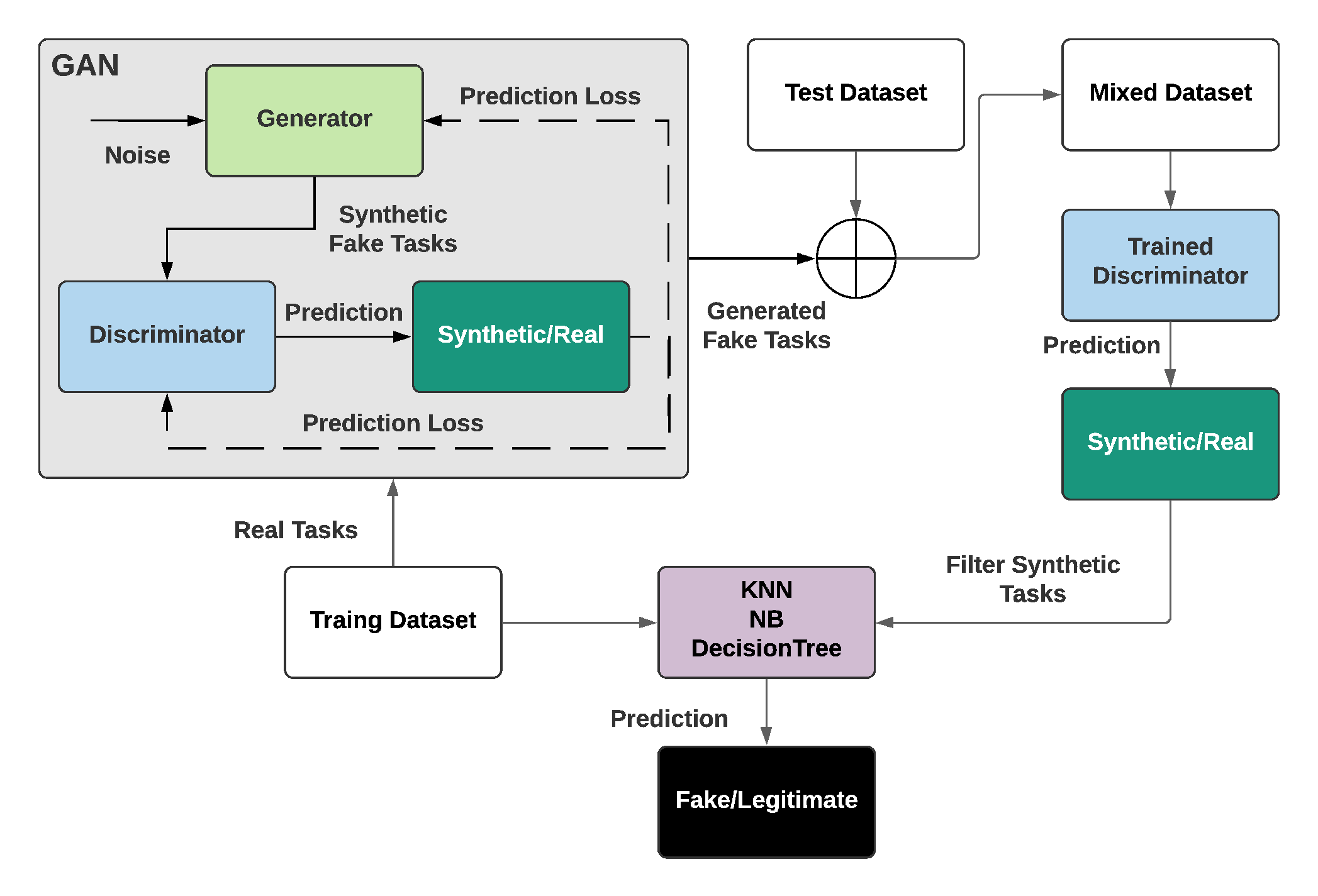}
    \caption{System of GAN-based adversarial tasks detection}
    \label{fig:MCS_system}

\end{figure}

\label{sec:AML}
We design a novel framework that implements a two-level cascaded classifier architecture to predict the generated attack samples and original (empirically designed) attack samples, and filter them before distributing them to MCS participants.  The first level is formed by the GAN discriminator whereas the second level is a binary classifier.
According to the system design in Fig.\ref{fig:MCS_system}, the training dataset is utilized to train binary classifiers (e.g., KNN, NB and DT) and the adversarial ML model, which is a GAN in this study. The training dataset includes the fake tasks of the adversaries in the first category, i.e, adversaries with low capacity (original fake tasks). GANs take input samples of noise, and output adversarial (i.e., synthetic) samples \cite{gan1}. An adversary with high capacity can design and train a GAN with its two neural network modules that are in competition with each other, namely the generator neural network and the discriminator neural network. The role of the generator is to create synthetic samples as similar as the real ones whereas a Discriminator designs to distinguish real samples from the synthetic ones. The motive for the competition is to reduce the gap between generated samples for Generator and increase detection accuracy for Discriminator \cite{wang2017generative}. 
On the basis of this motivation, the Generator and Discriminator participate in a non-cooperative game to reach a Nash Equilibrium~\cite{gan_nash_equil}.

\begin{algorithm}
\caption{Tasks in Mixed Dataset prediction procedure by Discriminator and classifier}
\label{alg:task_predic_proc}
\centering
\begin{algorithmic}[1]
\State $OriginalDataset$ = \{Fake\ Sensing\ Tasks\ by\ low-capacity\ adversaries\}
\State $GANDataset$ = \{Fake\ Sensing\ Tasks\ by\ high-\ capacity\ adversaries\}
\State $MixedDataset$ = $OriginalDataset \cup\ GANDataset$

\State $Label(Task_i)$ = 1, $\forall$($i|\ i\in MixedDataset$ $\wedge$ i: Original Task)
\State $Label(Task_i)$ = 0, $\forall$($i|\ i\in MixedDataset$ $\wedge$ i: Synthetic Task)
\Begin
\ForEach {$Task_i \in MixedDataset $}
\State $DP_{i} = Prediction\ of\ Discriminator\ for\ Task_{i}\ $
\If{$DP_{i} = 1$} 
 \Comment{ $Task_i$ is original (not synthetic). } \newline
 \Comment{Either legitimate (1) or illegitimate (0)}
 \State {Send\ $Task_i$\ to\ binary\ classifier}
    \State $CP_{i} =  Prediction\ of\ binary\ classifier\ for\ Task_{i}$
    \If{$CP_{i} = 1$}\newline
      \Comment{Classified as legitimate}
        \State{Accept\ sensing\ task\ request,\ pursue\ recruit\ participants}
    \Else
        \State{Eliminate\ $Task_{i}$\ By\ binary\ classifier}
    \EndIf
\Else
    \State Eliminate\ $Task_{i}$\ by\ Discriminator
    \EndIf
\End
\end{algorithmic}
\end{algorithm}

The generated fake tasks are injected to the MCS platform as sensing service requests. As the MCS platform is equipped with a Machine Learning (ML)-based decision system, handling of the synthetic tasks is interpreted from the standpoint of ML. In terms of ML, this means that the test dataset is injected synthetic fake tasks so these samples mix with the fake tasks of the adversaries with low capacity (i.e., original / empirically-designed fake tasks). Samples in this mixed dataset have two clusters, including real tasks in the test dataset (adversaries with low capacity, original fake tasks and legitimate tasks) and synthetic tasks (adversaries with high capacity, adversarial fake tasks). The Discriminator receives this mixed dataset, and it is already trained during the fake task creation procedure through the Generator of the GAN to distinguish synthetic from real samples so it can classify the two clusters in the mixed dataset. The Discriminator runs an estimation for the samples in the mixed dataset and filters out the samples predicted as 'synthetic fake task' before sending them to the level-2 classifier (e.g, KNN, NB, DT). In our previous work \cite{chencisose2022}, we showed that the ML-backed task detection module of the MCS platform is challenged by the adversarial fake tasks of the generator so in this paper, the Discriminator is utilized with motivation to improve the detection rate of the adversarial (synthetic) fake tasks. 
The rationale for bridging these two levels of prediction is as follows: It is possible that the Discriminator could predict a synthetic task as real, which will still be sent to the ML classifiers for prediction. 

Following upon the prediction of the Discriminator, the sensing tasks that are estimated as 'real' are picked. An real instance can have two classes: original (i.e., empirically designed) fake sensing tasks and legitimate sensing tasks. As mentioned earlier, the Discriminator is trained to detect and filter out the adversarial (GAN-generated) fake tasks. Thus, fake tasks attacks of the high capacity adversaries generated through a generative model are expected to be eliminated prior to the low capacity ML classifiers such as KNN, NB, or DT. As a result, these classifiers are expected to detect and eliminate the fake tasks submitted by the low capacity adversaries that set the task features through empirical analysis. Indeed, this two-level ML-backed fake sensing task detection system in the MCS platform aims to eliminate all fake tasks to maintain the MCS services available without the platform being clogged. Algorithm \ref{alg:task_predic_proc} describes the two-step prediction procedure for a sensing task in the mixed MCS dataset through the Discriminator and binary classifiers.

\section{Numerical Results and Analyses}
\label{sec:results}
This section presents the evaluation results and analyses regarding the impact of defensive use of the two-stage fake task detection to filter out the adversarial (submitted by high capacity adversaries) and original (submitted by low capacity adversaries) fake tasks in the MCS platforms.

\subsection{Dataset introduction and Assumptions}
The dataset used in this study is generated via a realistic MCS simulator, namely CrowdSenSim \cite{Crowdsensim}. We use realistic settings in Crowdsensim to simulate an MCS platform. The simulation configuration is shown in Table \ref{tab:crowdsensim_setup}, and adopted from \cite{zhang2019ai}. Clarence-Rockland is set as the city to conduct crowdsensing activities. A total of $14,484$ tasks have been generated under Crowdsensim with a breakdown of $12,587$ legitimate and $1,897$ illegitimate (fake) tasks. Train:test split of the dataset is set at 80\%:20\%. Legitimate and "original" fake tasks are generated based on the features/settings in the table. Hereafter, we refer to the fake tasks generated by the adversaries with low capacity (i.e., fake tasks that are empirically generated) as the original fake tasks. 

The rationale for the configuration of the fake task features in the simulation environment is that fake tasks aim to drain more resources from sensing service providing devices. Moreover, fake tasks seek longer sensing duration and higher volume sensing data as the ultimate goal is to clog the MCS sensing servers so the MCS platform and the sensing devices become unavailable. The feature set is adopted from \cite{zhang2019ai} to extract an MCS task with the features listed as follows: \{'ID', 'latitude', 'longitude', 'day', 'hour', 'minute', 'duration', 'remaining time', 'battery requirement (percentage)', 'Coverage', 'legitimacy', 'GridNumber', 'OnpeakHour'\}. 

\begin{figure}[ht]
    \centering
    \includegraphics[width = 0.8\linewidth, trim=3cm 0cm 3cm 1.5cm,clip]{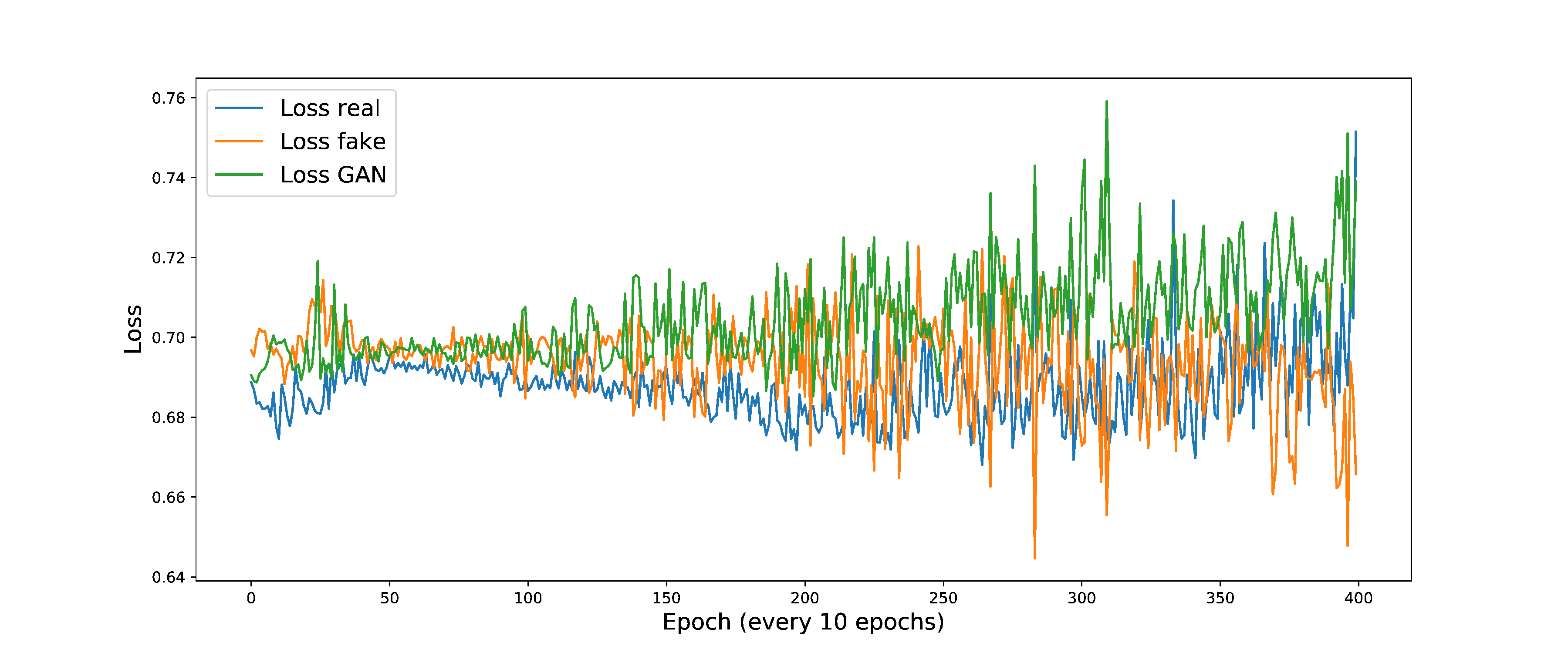}
    \caption{One round training loss for GAN}
    \label{fig:train_loss}

\end{figure}
\begin{table}[ht]
    \caption{ CrowdSenSin simulation setup for low capacity adversaries (adopted from \cite{zhang2019ai}).}
    \centering
    \label{tab:crowdsensim_setup}
\begin{tabular}{p{2.2cm}|p{2.3cm}|p{2.5cm}}
    \hline
    {}& Fake Tasks&Legitimate Tasks\\   
    \hline
    Day&\multicolumn{2}{c}{Uniformly distributed in [1, 6]}\\
    \hline
    Hour (task)&80\% : 7am to 11am;
    20\%: 12pm to 5pm& 8\%: 0pm to 5am; 92\%: 6am to 23pm\\
    \hline
    Duration (task)&70\% in \{40, 50, 60\} [minutes]; 
    30\% in \{10, 20, 30\} [minutes]&Uniformly distributed over \{10, 20, 30, 40, 50, 60\} mins\\
    \hline
    Battery usage (\%)&80\% in \{7\%-10\%\};
    20\% in \{1\% -6\%\}& Uniformly distributed in \{1\%-10\%\}\\
    \hline
    Recruitment Radius &\multicolumn{2}{c}{Uniformly distributed in 30m to 100m} \\
    \hline
    Movement Radius & \multicolumn{2}{c}{[10m, 80m]}\\
    \hline
    Number of tasks & \multicolumn{2}{c}{14,484}\\\hline
    \end{tabular}
\end{table}

\subsection{Adversarial task detection performance by Discriminator}
As illustrated in Fig \ref{fig:MCS_system}, the GAN model is trained with the training dataset so to have the adversaries with high capacity create fake sensing requests (i.e., tasks) via its Generator network. Discriminator and Generator are trained separately in batches. 

\begin{table}[]
\caption{GAN model hyper-parameters. G:Generator, D:Discriminator, LR:Learning Rate, AF: Activation Function}
\centering
\label{tab:gan_hyper}
\begin{tabular}{|c|c|c|c|c|c|}
\hline
& LR&\# of Neurons & AF   & BatchSize       & \# of Epochs    \\ \hline
G& 0.01 &256/512/1024 & tanh & 15,20,25,32 & 2k,4k,8k \\ \hline
D& 0.01&512/256/256 & sigmoid & 15,20,25,32 & 2k,4k,8k \\ \hline
\end{tabular}
\end{table}

As introduced before, dataset is split into training and test sets with a split ratio of 80\%-20\%. Training set contains  $1,506$ fake sensing tasks and $10,081$ legitimate tasks whereas the test set contains $391$ fake sensing tasks and $2,506$ legitimate tasks. It takes $2,000$ epochs for GAN to train itself. In order to demonstrate the reliability of test results, we repeat each test 20 times and present the average below in the results. TABLE \ref{tab:gan_hyper} demonstrates hyperparameters for GAN model. Specifically, in the Generator network, three layers are constructed with $256$, $512$ and $1024$ neurons respectively. Meanwhile, the activation function is selected as 'tanh'. On the other hand, in the Discriminator network, there are $512$, $256$, and $256$ neurons in each layer and the activation function is 'sigmoid'. Meanwhile, different batch sizes are applied to train the GAN model such as 15, 20, 32, and 64 in hyperparameter procedure. Finally, batch size of 32 is determined. Fig.\ref{fig:train_loss} demonstrates one round training loss for GAN, which presents the loss for every 10 epochs. It shows "loss real" for training loss for the Discriminator using real samples in the training dataset, "loss fake" for the training loss for the Discriminator using the generated samples by the Generator, and finally "loss GAN". In Fig. \ref{fig:train_loss}, network convergence is observed in $2000$ epochs. When training procedure completes, the Generator is used to generate a total of $2,000$ fake sensing requests. These $2,000$ fake tasks are designed by the adversaries with high capacity, and they are combined with the original fake sensing tasks in the test dataset to form the mixed dataset. It means there are two classes in the mixed dataset. We label real tasks in test dataset as class $1$ for both legitimate and fake tasks, and synthetic tasks as class $0$. 

\begin{table}[ht]
    \caption{Mixed dataset sample distribution}
    \centering

    \label{tab:Mix_DT_Distribution}
\begin{tabular}{|c|c|c|}
\hline
               & Class 0 (Adversarial)       & Class 1 (Real)      \\ \hline
Num of Fake Task           & 2000          & 391           \\ \hline
Num of Legitimate  Task   & 0             & 2506          \\ \hline
\textbf{Total} & \textbf{2000} & \textbf{2897} \\ \hline
\end{tabular}
\end{table}

The mixed dataset contains $2,897$ real tasks in test dataset (e.g., $391$ original fake tasks and $2,506$ legitimate tasks) and $2,000$ adversarial fake tasks, as shown in TABLE \ref{tab:Mix_DT_Distribution}. The discriminator pursues binary classification to distinguish real tasks from adversarial tasks. TABLE \ref{tab:Disc_confusionmatrix} presents the prediction confusion matrix, which represents the average of 20-round results. The Discriminator filters out samples that are predicted as adversaries, including the estimated real as adversaries (False Negative (FN)) and predicted as adversaries as adversaries(True Negative (TN)). On the other hand, the samples are predicted as real, including the prediction of real as real (True Positive (TP)) and the prediction of adversarial as real (False Positive (FP)) samples. Column \textit{IsFilterOut} in TABLE \ref{tab:Disc_confusionmatrix} denotes whether these samples would be removed or sent to the next-level classifier.

\begin{table}[ht]
    \caption{Distribution of the Discriminator predictions. Adver. means Adversarial.}
    \centering

    \label{tab:Disc_confusionmatrix}
\begin{tabular}{|c|c|c|c|}
\hline
Prediction                                                                                 & Task Type & Number of Tasks & \textit{IsFilterOut} \\ \hline
\multirow{2}{1.8cm}{\begin{tabular}[c]{@{}c@{}}Real as Real\end{tabular}}       & Original fake      & 175.1  & No          \\ \cline{2-4} 
                                                                                           & Legitimate & 2091.1 & No          \\ \hline
\multirow{2}{1.8cm}{\begin{tabular}[c]{@{}c@{}}Adver. as Real\end{tabular}}    & Adversarial fake      & 50.7   & No          \\ \cline{2-4} 
                                                                                           & Legitimate & N/A    & No          \\ \hline
\multirow{2}{1.8cm}{\begin{tabular}[c]{@{}c@{}}Real as Adver.\end{tabular}}    & Original fake      & 215.9  & Yes         \\ \cline{2-4} 
                                                                                           & Legitimate & 414.9  & Yes         \\ \hline
\multirow{2}{1.8cm}{\begin{tabular}[c]{@{}c@{}}Adver. as Adver.\end{tabular}} & Adversarial fake      & 1949.3 & Yes         \\ \cline{2-4} 
                                                                                           & Legitimate & N/A    & Yes         \\ \hline
\end{tabular}
\end{table}

\begin{table}[ht]
    \caption{Percentage of sensing tasks eliminated by the Discriminator due to predicting as adversaries}
    \centering
    \label{tab:Disc_elimi}
\begin{tabular}{|c|c|c|c|}
\hline
          & Num of prediction & Total & \%     \\ \hline
Original Fake      & 215.9    &391        & 55.2\% \\ \hline
Legitimate & 414.9     &2506       & 16.6\% \\ \hline
Adversarial Fake & 1949.3   &2000        & 97.5\% \\ \hline
\end{tabular}
\end{table}

As seen in TABLE \ref{tab:Disc_confusionmatrix}, the Discriminator predicts $50.7$ adversarial fake tasks as real tasks, which are marked to be sent to the next level classifier (KNN/NB/DT) to determine make a final decision of them being fake or legitimate. Meanwhile, as seen in the FN breakdown, out of $630.8$ real tasks, $215.9$ fake sensing requests and $414.9$ legitimate sensing requests are predicted as synthetic ones, which are eliminated before being sent to the next-level classifier. Table \ref{tab:Disc_elimi} illustrates the elimination percentage for real tasks and adversarial fake tasks. The Discriminator detection rate for synthetic samples is up to $97.5\%$, which means that most generated samples by the adversaries with high capacity are prevented in the MCS systems. Moreover, the Discriminator marks $55.2\%$ original fake tasks as synthetic tasks and excludes them. However, $16.6\%$ ($414.9$) legitimate tasks are estimated as synthetic ones and eradicated. Therefore, there is trade-off between fake tasks detection and legitimate tasks completion.

\begin{table}[ht]
    \caption{KNN, NB, and DT correct prediction of fake sensing tasks with/without Discriminator}
    \centering
    \label{tab:KNN_NB_DT_TN}
\begin{tabular}{|c|c|c|c|c|}
\hline
HasDiscriminator      & FakeTaskFrom    & KNN   & NB   & DT    \\ \hline
\multirow{2}{1.8cm}{No}  & Original    & 299   & 102  & 391   \\ \cline{2-5} 
                     & Adversarial & 0     & 0    & 918.8     \\ \hline
\multirow{2}{1.8cm}{Yes} & Original    & 129.5 & 24.5 & 175.1 \\ \cline{2-5} 
                     & Adversarial & 0     & 0    & 50.7  \\ \hline
\end{tabular}
\end{table}

\subsection{Adversarial task detection performance by the second level classifiers}
Following upon the Discriminator's filtering, the remaining samples are fed to the next-level (i.e., binary) classifiers, which are trained with the training dataset. The distribution of the remaining samples is presented in Table \ref{tab:Disc_confusionmatrix} for the TP and FP prediction samples. There are $2,316.9$ tasks in total, with $225.8$ fake tasks ($175.1$ original fake tasks and $50.7$ adversarial fake tasks) and $2,091.1$ legitimate tasks.

Three classifiers are selected for the fake task detection in the MCS platform, including Naive Bayes (NB), Decision Tree (DT), and K-Nearest Neighbors (KNN) to predict whether a sensing task is legitimate or fake. In order to present the effectiveness of the proposed architecture, we compare the performance of the two-level architecture to the architecture without the Discriminator, which stands for the case where the mixed dataset in TABLE \ref{tab:Mix_DT_Distribution} is directly fed to the binary classification. 

Three metrics, including Adversarial Attack Success Rate (AASR), Adversarial Attack Detection Rate (AADR), and Original Attack Detection Rate (OADR), are used to evaluate the  system performance. AADR (\ref{equ:AADR}) represents the ability of a defense system to prevent adversarial attacks, which is the ratio of detected adversarial attacks to the total adversarial instances \cite{ids_tech9} .
\begin{equation}
    \label{equ:AADR}
    AADR = \frac{DA\_DIS+DA\_CLA}{Total Adversarial}
\end{equation}
In Eq. \ref{equ:AADR}, $DA\_DIS$ denotes the number of detected adversarial attack samples by the Discriminator, and $DA\_CLA$ means the number of detected adversarial attack samples by a classifier (e.g., KNN,NB,DT). Since our simulations generate $2,000$ fake tasks by the Generator, the value of $Total Adversarial$ is $2,000$.

AASR in represents the adversarial attack successful injection rate. Sum of AADR and AASR (\(AASR+AADR = 1\)) is $100\%$, which means an adversarial attack either be detected or injected successfully to an MCS system.

Original Attack Detection Rate (OADR) shows the effectiveness for the detection system in identifying the original fake sensing requests submitted by adversaries with low capacity. The formulation shows in (\ref{equ:OADR}).
\begin{equation}
    \label{equ:OADR}
    OADR = \frac{DO\_DIS + DO\_CLA} {Total\_Original\_Attacks}
\end{equation}

In Eq. \ref{equ:OADR}, $DO\_DIS$ represents the number of detected original fake tasks by the Discriminator and $DO\_CLA$ denotes the number of detected original fake tasks by a classifier (e.g., KNN,NB,DT). As introduced before, there are $391$ fake sensing tasks in the test dataset so $Total\_Original\_Attacks$ is $391$.

\begin{table}[hb]
    \caption{With/Without Discriminator results comparison for AASR, AADR, and OADR with/without Discriminator}
    \centering

    \label{tab:ASR_comp}
    \begin{tabular}{|c|c|c|c|c|c|}
    \hline
                          & HasDiscr             & Contributed By   & KNN              & NB               & DT               \\ \hline
    \multirow{4}{0.6cm}{AASR} & No                   & Classifier       & 1.000           & 1.000           & 0.459           \\ \cline{2-6} 
                          & \multirow{3}{0.6cm}{Yes} & Discriminator    & 0.025          & 0.025          & 0.025          \\ \cline{3-6} 
                          &                      & Classifier       & 0.025          & 0.025          & 0                \\ \cline{3-6} 
                          &                      & \textbf{Finally} & \textbf{0.025} & \textbf{0.025} & \textbf{0}       \\ \hline
    \multirow{4}{0.6cm}{AADR} & No                   & Classifier       & 0.000            & 0.000            & 0.459            \\ \cline{2-6} 
                          & \multirow{3}{0.6cm}{Yes} & Discriminator    & 0.975            & 0.975            & 0.975            \\ \cline{3-6} 
                          &                      & Classifier       & 0.975            & 0.975            & 1.000            \\ \cline{3-6} 
                          &                      & \textbf{Finally} & \textbf{0.975}   & \textbf{0.975}   & \textbf{1.000}   \\ \hline
    \multirow{4}{0.6cm}{OADR} & No                   & Classifier       & 0.765            & 0.261            & 1.000            \\ \cline{2-6} 
                          & \multirow{3}{0.6cm}{Yes} & Discriminator    & 0.552            & 0.552            & 0.552            \\ \cline{3-6} 
                          &                      & Classifier       & 0.331            & 0.063            & 0.448            \\ \cline{3-6} 
                          &     & \textbf{Finally}   & \textbf{0.883}  & \textbf{0.615}  & \textbf{1.000} \\ \hline
    \end{tabular}
\end{table}

TABLE \ref{tab:KNN_NB_DT_TN} demonstrates the number of correct predictions for fake tasks by classifiers (e.g., KNN, NB and DT), which contains original fake sensing requests and fake sensing requests designed by high capacity adversaries. As seen in the table, KNN and NB cannot detect any adversarial tasks regardless of the presence of the Discriminator. DT can recognize adversarial fake tasks under both scenarios (i.e., with or without the Discriminator). It is worth to note that with the Discriminator, $50.7$ original fake tasks are identified correctly by DT, which is much lower than $918.8$ without the Discriminator case. This is because the Discriminator eliminates $215.9$ original fake tasks as shown in Table \ref{tab:Disc_confusionmatrix}. According to the prediction results by the Discriminator and classifiers, as well as AASR, AADR, and OADR metrics (see Eq. \ref{equ:AADR} and \ref{equ:OADR}), we calculate these three metrics.

AASR, AADR, and OADR results are shown in TABLE \ref{tab:ASR_comp}. It is evident that the architecture with the Discriminator results improves performance when compared to the architecture without the Discriminator. For instance, KNN and NB perform $100\%$ AASR without the Discriminator, that denotes KNN and NB have no ability to prevent adversarial attacks. On the other hand, AASR is reduced significantly to $2.5\%$ by using Discriminator in TABLE \ref{tab:ASR_comp}. Moreover, DT identifies $2.5\%$ of the adversarial fake tasks, resulting in $0\%$ AASR finally. Regarding to AADR, it can be calculated according to AASR in Eq. (\ref{equ:AADR}). Moreover, a lower AASR leads to a higher AADR in TABLE \ref{tab:ASR_comp}. Original fake tasks are filtered by the Discriminator firstly, and then by binary classifiers so the finally OADR is the sum of OADR by the Discriminator and OADR by the second-level classifier. DT results in the best detection performance up $100\%$ accuracy for both with and without Discriminator cases. The original fake task detection rate is improved, when with and without the Discriminator architectures are compared. The improvement is from $76.5\%$ to $88.3\%$ under KNN and from $26.1\%$ to $61.5\%$ under NB.

\section{Conclusion and Open Issues}
\label{sec:conclusion}
 
Adversaries with high capacity have the potential to integrate intelligent techniques to design attacks on Mobile Crowdsensing (MCS) systems, which can lead to more damage than normal attacks due to the low detection rate via traditional ML-based classifiers. 
In this paper, we have adopted the previous work to generate attack samples via GAN and we propose a novel two-level framework to integrate the Discriminator of a Generative Adversarial Network (GAN) and low-capacity binary classifiers.  
Binary classifiers (e.g., KNN, Naive Bayes (NB), Decision Tree (DT)) are used to estimate whether a sensing task is fake or legitimate. 
Numerical results illustrate that the proposed approach boosts the Adversarial Attack Detection Rate (AADR) when compared to the detection model without the Discriminator. Thus, detection performance of KNN and NB increase from $0$ to $97.5\%$, respectively whereas DT raises from $45.9\%$ to $100\%$. Thus, all generated adversarial fake sensing requests are identified by the DT in the second level, and AASR is $0$. Meanwhile, the proposed method increases the performance for the original fake sensing tasks, with OADR boosting from $76.5\%$ to $88.3\%$ under KNN, $26.1\%$ to $61.5\%$ under NB. DT achieves $100\%$ in terms of OADR, which means all original fake tasks are filtered out. DT achieves the best performance than NB and DT. 

Indeed, maintaining the effectiveness of an MCS system while eliminating the fake sensing requests by high capacity adversaries remains an open issue, which will be tackled by our ongoing research agenda.

\bibliographystyle{ieeetr}
\bibliography{refs.bib}

\end{document}